\documentclass[11pt]{article}

\usepackage{amsmath}
\usepackage{graphicx}
\usepackage{rotating}
\usepackage{multirow}
\usepackage{tikz}
\usepackage{pgfplots}
\usepackage{tabulary}
\usepackage{tabu}
\usepackage{colortbl}
\usetikzlibrary{matrix,positioning,chains}

\usepackage[a4paper, total={6.5in, 9in}]{geometry}
\usepackage[font=small]{caption}

\tikzstyle{phase1}=[
  rectangle,
  rounded corners,
  minimum width=1cm,
  text width=3cm,
  text centered,
  minimum height=0.6cm,
  draw=black,
]

\tikzstyle{phase2}=[
  rectangle,
  rounded corners,
  minimum width=1cm,
  text width=3cm,
  text centered,
  minimum height=0.6cm,
  draw=black,
]

\begin{document}

\title{Accurate selfcorrection of errors in long reads\\using de Bruijn graphs}

\author{Leena Salmela$^1$ \and Riku Walve$^1$ \and Eric Rivals$^2$ \and Esko Ukkonen$^1$}
\date{\small $^1$Helsinki Institute for Information Technology HIIT,\\
Department of Computer Science, University of Helsinki\\
\{lmsalmel,riqwalve,ukkonen\}@cs.helsinki.fi\\
$^2$LIRMM and Institut de Biologie Computationelle, \\
CNRS and Universit\'e Montpellier, Montpellier, France\\
rivals@lirmm.fr
}

\maketitle

\begin{abstract}
New long read sequencing technologies, like PacBio SMRT and Oxford NanoPore, can produce sequencing reads up to 50,000 bp long but with an error rate of at least 15\%. Reducing the error rate is necessary for subsequent utilisation of the reads in, e.g., {\em de novo} genome assembly. The error correction problem has been tackled either by aligning the long reads against each other or by a hybrid approach that uses the more accurate short reads produced by second generation sequencing technologies to correct the long reads. We present an error correction method that uses long reads only. The method consists of two phases: first we use an iterative alignment-free correction method based on de Bruijn graphs with increasing length of $k$-mers, and second, the corrected reads are further polished using long-distance dependencies that are found using multiple alignments. According to our experiments the proposed method is the most accurate one relying on long reads only for read sets with high coverage. Furthermore, when the coverage of the read set is at least 75x, the throughput of the new method is at least 20\% higher. LoRMA is freely available at {http://www.cs.helsinki.fi/u/lmsalmel/LoRMA/}.
\end{abstract}


\section{Introduction}

With the diminishing costs, high throughput DNA sequencing has become a commonplace technology in biological research. Whereas the second generation sequencers produced short but quite accurate reads, new technologies such as Pacific Biosciences and Oxford NanoPore are producing reads up to 50,000 bp long but with an error rate at least 15\%. 
Although the long reads have proven to be very helpful in applications like genome assembly~\cite{koph14,nanoassembly}, the error rate poses a challenge for the utilisation of this data. 

Many methods have been developed for correcting 
short reads~\cite{ectool,laehnemann15} but these methods are not directly applicable to the long reads because of their much higher error rate. Moreover, most research of short read error correction has concentrated on mismatches, the dominant error type in Illumina data, whereas in long reads indels are more common. Recently several methods for error correction of long reads have also been developed. These methods fall into two categories: either the highly erroneous long reads are selfcorrected by aligning them against each other, or a hybrid strategy is adopted in which 
the long reads are corrected using the accurate short 
reads that are assumed to be available.
Most standalone error correction tools like
proovread~\cite{proovread}, LoRDEC~\cite{lordec}, LSC~\cite{lsc}, and Jabba~\cite{jabba}
are hybrid methods. PBcR~\cite{pacbiotoca,mhap} is a tool that can
employ either the hybrid or selfcorrection strategy.

Most hybrid methods like PBcR, LSC, and proovread are based on the mapping approach. 
They first map the short reads on the long reads and then correct the long reads according to a consensus 
built on the mapped short reads. PBcR extends this strategy to selfcorrection of PacBio reads 
by computing overlaps between the long reads using probabilistic locality-sensitive hashing and then correcting the reads according to a consensus built 
on the overlapping reads. As the mapping of short reads is time and memory consuming, LoRDEC avoids the 
mapping phase by building a de Bruijn graph (DBG) of the short reads and then threading the long reads 
through this graph to correct them. Jabba is a recent tool that is also based on building a DBG of short reads. 
While LoRDEC finds matches of complete $k$-mers in the long reads, Jabba searches for maximal exact matches 
between the $k$-mers and the long reads allowing it to use a larger $k$ in the DBG.

In this paper, we present a selfcorrection method for long reads that is based on de Bruijn graphs and multiple alignments. 
First our method performs initial correction that is similar to LoRDEC, but uses only long reads and performs iterative correction rounds with longer and longer $k$-mers. 
This phase considers only the local context of errors and hence it misses the long-distance dependency information available in the long reads.
To capture such dependencies, the second phase of our method uses multiple alignments between 
carefully selected reads to further improve the error correction.

Our experiments show that our method is currently the most accurate one relying on long reads only. The error rate of the reads after our error correction is less than half of the error rate of reads corrected by PBcR using long reads only.
Furthermore, when the coverage of the read set is at least 75x, the size of the corrected read set of our method is at least 20\% higher than for PBcR.

\section{Overview of LoRDEC}

LoRDEC~\cite{lordec} is a hybrid method for the error correction of long reads. 
It 
presents the short reads in a de Bruijn graph (DBG) and then maps the long reads to the graph. The DBG of a read set is a graph whose nodes are all $k$-mers occurring in the reads and there is an edge between two nodes if the corresponding $k$-mers overlap by $k-1$ bases.
LoRDEC classifies the $k$-mers of long reads as {\em solid} if they are in the DBG and {\em weak} otherwise. The correction then proceeds by replacing the weak areas of the long reads by solid ones. This is done by searching paths in the DBG between solid $k$-mers to bridge the weak areas between them. If several paths are found, the path with the shortest edit distance as compared to the weak region is chosen to be the correct sequence, which replaces the weak region of the long read. The weak heads and tails of the long reads are the extreme regions of the reads that are bordered by just one solid $k$-mer in the beginning (resp. end) of the read. LoRDEC attempts to correct these regions by starting a path search from the solid $k$-mer and choosing a sequence that is as close as possible to the weak head or tail.

Repetitive regions of the genome can make the DBG tangled. The path search in these areas of the DBG can then become intractable. Therefore LoRDEC employs a limit on the number of branches it explores during the search. If this limit is exceeded, LoRDEC checks if at least one path within the maximum allowed error rate has been found and then uses the best path found for correction. If no such path has been found, LoRDEC starts a path search similar to the correction of the head and tail of the read, to attempt a partial correction of the weak region.

Some segments of the long reads remain erroneous after the correction. LoRDEC outputs bases in upper case if at least one of the $k$-mers containing that base is solid, i.e., it occurs in the DBG of the short reads, and in lower case otherwise. For most applications it is preferable to extract only the upper case regions of the sequences as the lower case bases are likely to contain errors.

\section{Selfcorrection of long reads}

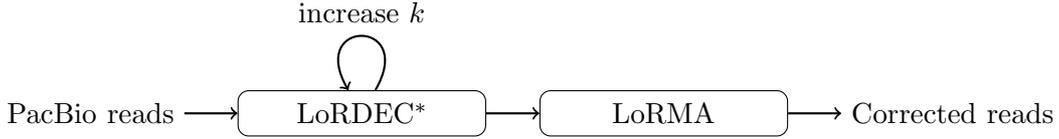
\begin{figure}
  \centering

  \begin{tikzpicture}
    \node (pbreads) {PacBio reads};
    \node[phase1, right = 0.7cm of pbreads] (LoRDEC) {LoRDEC$^*$};
    \node[phase2, right = 0.7cm of LoRDEC] (LoRMA) {LoRMA};
    \node[right = 0.7cm of LoRMA] (creads) {Corrected reads};

    \draw[->, thick] (pbreads) -- (LoRDEC);
    \draw[phase2, ->, thick] (LoRDEC) to [out=60, in=120, looseness=8] node[above] {increase $k$} (LoRDEC);
    \draw[->, thick] (LoRDEC) -- (LoRMA);
    \draw[->, thick] (LoRMA) -- (creads);
  \end{tikzpicture}
  \caption{Workflow of error correction. LoRDEC$^*$ is first applied iteratively to the read set, with an increasing $k$. The corrected reads are further corrected by LoRMA which uses multiple alignments 
to find long-distance dependencies in the reads.}
  \label{fig:workflow}
\end{figure}

In this section we will show how an error correction procedure similar to LoRDEC can be used to iteratively correct long reads without short read data. We will use LoRDEC$^*$ to refer to LoRDEC in this long reads only mode.
Then we further describe a polishing method to improve the accuracy
of correction. Figure~\ref{fig:workflow} shows the workflow of our
approach.

\subsection{Iterative correction}

To describe how LoRDEC can be adapted for selfcorrection of read sets, let $Q$ be a set of long reads to be corrected, and
let integer $h$ be the {\em abundancy threshold}
that is used in choosing the $k$-mers to the DBG. 
The correction procedure repeats for 
an increasing sequence $k = k_1, \ldots, k_t$ the following steps 1--3: 
\begin{enumerate}
\item Construct the DBG of set $Q$ using 
as the nodes
the $k$-mers that occur in $Q$ at least $h$ times; 
\item Correct $Q$ using the LoRDEC algorithm with this DBG; 
\item Replace $Q$ with the corrected $Q$.
\end{enumerate}
After the final round, the regions of the reads identified as correct in the last iteration
are extracted for further 
correction with the multiple alignment technique by LoRMA.


As the initial error level is assumed high, 
the above iterations have to start with a relatively small $k=k_1$. 
With a suitable abundancy threshold $h$, the DBG should then contain most of the correct 
$k$-mers (i.e., the $k$-mers of the target genome)
and a few erroneous ones. 
Although path search
over long weak regions may not be feasible because of strong branching of the DBG, 
shorter paths are likely to be found and hence,
short weak regions can be corrected. 
After the first round the
correct regions in the reads have become longer because close-by correct
regions have been merged whenever a path between 
them has been found, and thus we can increase $k$.
Then, with increasing $k$s, the DBG gets less tangled and the path search over the longer
weak regions becomes feasible allowing for the correction of the
complete reads.
A similar iterative approach has previously been proposed for short read
assembly \cite{idba,spades}.

When the path search is abandoned because of excessive branching, 
the original
LoRDEC algorithm still uses the best path found so far to correct the region. 
Such a
greedy
strategy improves correction accuracy in a single run, 
but in the present iterative approach
false
corrections start to accumulate.
Therefore, we make a correction only
if it is guaranteed that the correction is the best one available in
the DBG, i.e., all branches have been explored.

Abundancy threshold $h$ controls the quality of the $k$-mers that are used for correction.
In our experiments we used a fixed threshold of $h=4$ in all iterations, meaning that the $k$-mers with less than $4$ occurrences 
in the read set were considered erroneous. 

To justify the value of $h$, 
we need to analyse how many times a fixed $k$-mer of the genome is expected to occur
without any error in the reads.
Then an $h$ that is about one or two standard deviations below the expected value should give a DBG
that contains the majority of the correct $k$-mers and not too many erroneous ones.
We will 
use an analysis similar to Miclotte et al.\cite{jabba}.

\begin{figure}
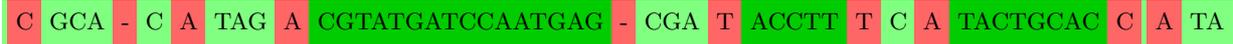

\begin{center}
{\footnotesize
\colorbox{green!50}{\strut\hspace{-0.14cm}}\colorbox{red!60}{\strut C}\colorbox{green!50}{\strut GCA}\colorbox{red!60}{\strut -}\colorbox{green!50}{\strut C}\colorbox{red!60}{\strut A}\colorbox{green!50}{\strut TAG}\colorbox{red!60}{\strut A}\colorbox{green!80!black}{\strut CGTATGATCCAATGAG}\colorbox{red!60}{\strut -}\colorbox{green!50}{\strut CGA}\colorbox{red!60}{\strut T}\colorbox{green!80!black}{\strut ACCTT}\colorbox{red!60}{\strut T}\colorbox{green!50}{\strut C}\colorbox{red!60}{\strut A}\colorbox{green!80!black}{\strut TACTGCAC}\colorbox{red!60}{\strut C}\colorbox{green!50}{\strut\hspace{-0.14cm}}\colorbox{red!60}{\strut A}\colorbox{green!50}{\strut TA}
}
\caption{
Division of a read into maximal exact regions, shown as green areas. 
The dark green areas give the regions that could cover a 4-mer. 
}
\label{fig:ecr-example}
\end{center}
\end{figure}

Let $C_{\ell \geq k}$ denote
{\em the coverage of a genomic $k$-mer by exact regions of length at least $k$}.
Here {\em exact region} refers to a continuous maximal error-free segment of some read 
in our read set.
Figure~\ref{fig:ecr-example} gives an example of exact regions.
Let us add a \$ character to the end of each read, and then consider the concatenation 
of all these 
reads. 
In this sequence an exact region (of length 0 or more) ends either at an error or 
when encountering the \$ character. 
Let $n$ denote the number of reads, $N$ the length of the concatenation of all reads, 
and $p$ the error rate. 
Then the probability for an exact region to end at a given position of 
the concatenated sequence is $q=(pN+n)/(N+n)$. 
As the reads are long and the error rate is high, we have $q\approx p$. 
The length of the exact regions is distributed according to the geometric distribution $\mathrm{Geom}(q)$ and 
therefore the probability of an exact region to have length $i$ is $P(i) = (1-q)^iq$. 
The expected number of exact regions is $Nq$. 
An exact region is {\em maximal} if it cannot be extended to the left or right.
Let $R_i$ be the random variable denoting 
the number of maximal exact regions of length $i$. 
Then $E(R_i) = N q P(i)=Nq^2(1-q)^i$.

Let $C_{\ell =i}$ denote the coverage of a $k$-mer in the genome by maximal exact regions of length $i$,
and let  
$r_i$ denote the number of maximal exact regions of length $i$. 
An exact region of length $i$, $i \geq k$, covers a fixed genomic $k$-mer
(i.e., the read with that exact region is read from the genomic segment containing that $k$-mer)
if the region starts in the genome from the starting location of the $k$-mer
or from some of the $i-k$ locations before it.
Assuming that the reads are randomly sampled from the genome, this happens with probability
$(i-k+1)/G$, where $G$ is the length of the genome.
Therefore, $C_{\ell=i}$ is distributed according to the binomial distribution $\mathrm{Bin}(r_i, (i-k+1)/G)$ 
(independence of locations of exact regions is assumed), and 
the expected coverage of a genomic $k$-mer by maximal exact regions of length $i$ is 
\begin{equation*}
\begin{split}
E(C_{\ell=i}) & = \sum_{r_i=0}^\infty P(R_i=r_i)\cdot r_i\cdot \frac{i-k+1}{G} \\
& = \frac{i-k+1}{G} E(R_i) \\
& = \frac{N}{G} q^2(1-q)^i\cdot (i-k+1).
\end{split}
\end{equation*}
By the linearity of expectation the expected coverage of a genomic $k$-mer by exact regions of length at least $k$ is
\begin{equation*}
\begin{split}
E(C_{\ell \geq k}) & = \sum_{i=k}^{\infty} E(C_{\ell =i}) \\
                & = \frac{N}{G}\sum_{i=k}^{\infty} q^2(1-q)^i\cdot (i-k+1) .
\end{split}
\end{equation*}

Because $(i-k+1)/G$ is small, we can approximate the binomial distribution of $C_{\ell =i}$ with the Poisson distribution. 
Therefore $\sigma^2(C_{\ell =i}) = E(C_{\ell =i})$. 

Assuming that
the coverages of a genomic $k$-mer by maximal exact regions of different lengths 
are
independent, 
the variance of the coverage 
by exact regions of length 
at least $k$ is $\sigma^2(C_{\ell \geq k}) = \sum_{i\geq k} \sigma^2(C_{\ell=i}) = E(C_{\ell \geq k})$.

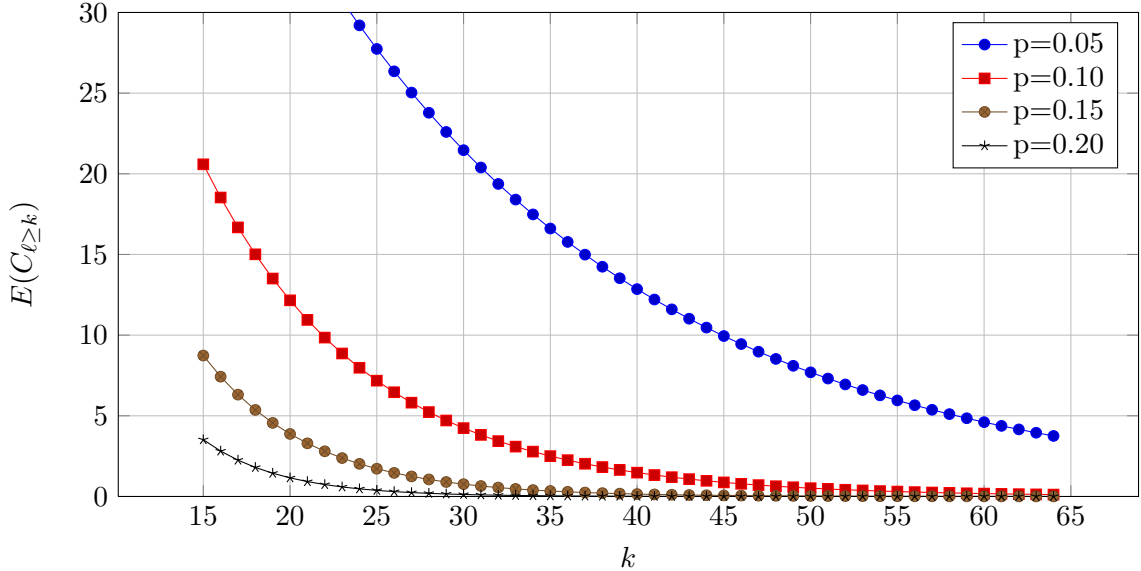
\begin{figure}
  \begin{center}

  \begin{tikzpicture}
	\begin{axis}[ 
          xlabel=$k$,
          ylabel={$E(C_{\ell \geq k})$},
          ymin=0,
          ymax=30,
          height=8cm, 
          width=15cm, 
          grid=major,
          tick label style={/pgf/number format/fixed}, ] 
          \addplot coordinates 
          { 
( 15 , 46.329123016 )
( 16 , 44.0126668652 )
( 17 , 41.8120335219 )
( 18 , 39.7214318458 )
( 19 , 37.7353602535 )
( 20 , 35.8485922409 )
( 21 , 34.0561626288 )
( 22 , 32.3533544974 )
( 23 , 30.7356867725 )
( 24 , 29.1989024339 )
( 25 , 27.7389573122 )
( 26 , 26.3520094466 )
( 27 , 25.0344089742 )
( 28 , 23.7826885255 )
( 29 , 22.5935540993 )
( 30 , 21.4638763943 )
( 31 , 20.3906825746 )
( 32 , 19.3711484459 )
( 33 , 18.4025910236 )
( 34 , 17.4824614724 )
( 35 , 16.6083383988 )
( 36 , 15.7779214788 )
( 37 , 14.9890254049 )
( 38 , 14.2395741346 )
( 39 , 13.5275954279 )
( 40 , 12.8512156565 )
( 41 , 12.2086548737 )
( 42 , 11.59822213 )
( 43 , 11.0183110235 )
( 44 , 10.4673954723 )
( 45 , 9.94402569871 )
( 46 , 9.44682441377 )
( 47 , 8.97448319309 )
( 48 , 8.52575903343 )
( 49 , 8.09947108176 )
( 50 , 7.69449752767 )
( 51 , 7.30977265129 )
( 52 , 6.94428401872 )
( 53 , 6.59706981779 )
( 54 , 6.2672163269 )
( 55 , 5.95385551055 )
( 56 , 5.65616273503 )
( 57 , 5.37335459827 )
( 58 , 5.10468686836 )
( 59 , 4.84945252494 )
( 60 , 4.6069798987 )
( 61 , 4.37663090376 )
( 62 , 4.15779935857 )
( 63 , 3.94990939064 )
( 64 , 3.75241392111 )
        };
          \addlegendentry{p=0.05};
          \addplot coordinates 
          { 
( 15 , 20.5891132095 )
( 16 , 18.5302018885 )
( 17 , 16.6771816997 )
( 18 , 15.0094635297 )
( 19 , 13.5085171767 )
( 20 , 12.1576654591 )
( 21 , 10.9418989132 )
( 22 , 9.84770902184 )
( 23 , 8.86293811965 )
( 24 , 7.97664430769 )
( 25 , 7.17897987692 )
( 26 , 6.46108188923 )
( 27 , 5.8149737003 )
( 28 , 5.23347633027 )
( 29 , 4.71012869725 )
( 30 , 4.23911582752 )
( 31 , 3.81520424477 )
( 32 , 3.43368382029 )
( 33 , 3.09031543826 )
( 34 , 2.78128389444 )
( 35 , 2.50315550499 )
( 36 , 2.25283995449 )
( 37 , 2.02755595904 )
( 38 , 1.82480036314 )
( 39 , 1.64232032683 )
( 40 , 1.47808829414 )
( 41 , 1.33027946473 )
( 42 , 1.19725151826 )
( 43 , 1.07752636643 )
( 44 , 0.969773729788 )
( 45 , 0.872796356809 )
( 46 , 0.785516721128 )
( 47 , 0.706965049015 )
( 48 , 0.636268544114 )
( 49 , 0.572641689702 )
( 50 , 0.515377520732 )
( 51 , 0.463839768659 )
( 52 , 0.417455791793 )
( 53 , 0.375710212614 )
( 54 , 0.338139191352 )
( 55 , 0.304325272217 )
( 56 , 0.273892744995 )
( 57 , 0.246503470496 )
( 58 , 0.221853123446 )
( 59 , 0.199667811102 )
( 60 , 0.179701029991 )
( 61 , 0.161730926992 )
( 62 , 0.145557834293 )
( 63 , 0.131002050864 )
( 64 , 0.117901845777 )
          };
          \addlegendentry{p=0.10};
          \addplot coordinates 
          {
( 15 , 8.73542191013 )
( 16 , 7.42510862361 )
( 17 , 6.31134233007 )
( 18 , 5.36464098056 )
( 19 , 4.55994483347 )
( 20 , 3.87595310845 )
( 21 , 3.29456014218 )
( 22 , 2.80037612086 )
( 23 , 2.38031970273 )
( 24 , 2.02327174732 )
( 25 , 1.71978098522 )
( 26 , 1.46181383744 )
( 27 , 1.24254176182 )
( 28 , 1.05616049755 )
( 29 , 0.897736422916 )
( 30 , 0.763075959479 )
( 31 , 0.648614565557 )
( 32 , 0.551322380724 )
( 33 , 0.468624023615 )
( 34 , 0.398330420073 )
( 35 , 0.338580857062 )
( 36 , 0.287793728503 )
( 37 , 0.244624669227 )
( 38 , 0.207930968843 )
( 39 , 0.176741323517 )
( 40 , 0.150230124989 )
( 41 , 0.127695606241 )
( 42 , 0.108541265305 )
( 43 , 0.092260075509 )
( 44 , 0.0784210641826 )
( 45 , 0.0666579045552 )
( 46 , 0.0566592188719 )
( 47 , 0.0481603360411 )
( 48 , 0.040936285635 )
( 49 , 0.0347958427897 )
( 50 , 0.0295764663713 )
( 51 , 0.0251399964156 )
( 52 , 0.0213689969532 )
( 53 , 0.0181636474103 )
( 54 , 0.0154391002987 )
( 55 , 0.0131232352539 )
( 56 , 0.0111547499658 )
( 57 , 0.00948153747095 )
( 58 , 0.00805930685031 )
( 59 , 0.00685041082276 )
( 60 , 0.00582284919935 )
( 61 , 0.00494942181945 )
( 62 , 0.00420700854653 )
( 63 , 0.00357595726455 )
( 64 , 0.00303956367487 )
          };
          \addlegendentry{p=0.15};
          \addplot coordinates 
          {
( 15 , 3.51843720888 )
( 16 , 2.81474976711 )
( 17 , 2.25179981369 )
( 18 , 1.80143985095 )
( 19 , 1.44115188076 )
( 20 , 1.15292150461 )
( 21 , 0.922337203685 )
( 22 , 0.737869762948 )
( 23 , 0.590295810359 )
( 24 , 0.472236648287 )
( 25 , 0.37778931863 )
( 26 , 0.302231454904 )
( 27 , 0.241785163923 )
( 28 , 0.193428131138 )
( 29 , 0.154742504911 )
( 30 , 0.123794003929 )
( 31 , 0.0990352031428 )
( 32 , 0.0792281625143 )
( 33 , 0.0633825300114 )
( 34 , 0.0507060240091 )
( 35 , 0.0405648192073 )
( 36 , 0.0324518553658 )
( 37 , 0.0259614842927 )
( 38 , 0.0207691874341 )
( 39 , 0.0166153499473 )
( 40 , 0.0132922799578 )
( 41 , 0.0106338239663 )
( 42 , 0.00850705917302 )
( 43 , 0.00680564733842 )
( 44 , 0.00544451787074 )
( 45 , 0.00435561429659 )
( 46 , 0.00348449143727 )
( 47 , 0.00278759314982 )
( 48 , 0.00223007451985 )
( 49 , 0.00178405961588 )
( 50 , 0.00142724769271 )
( 51 , 0.00114179815416 )
( 52 , 0.000913438523332 )
( 53 , 0.000730750818665 )
( 54 , 0.000584600654932 )
( 55 , 0.000467680523946 )
( 56 , 0.000374144419157 )
( 57 , 0.000299315535325 )
( 58 , 0.00023945242826 )
( 59 , 0.000191561942608 )
( 60 , 0.000153249554087 )
( 61 , 0.000122599643269 )
( 62 , 9.80797146154e-05 )
( 63 , 7.84637716923e-05 )
( 64 , 6.27710173539e-05 )
          };
          \addlegendentry{p=0.20};
        \end{axis}

  \end{tikzpicture}

  \caption{Expected coverage of a genomic $k$-mer by exact regions of length at least $k$ 
for a read set with coverage 100x for different error rates $p$.}
  \label{fig:ecr}
  \end{center}
\end{figure}

Figure~\ref{fig:ecr} 
illustrates $E(C_{\ell \geq k})$ for various $k$ and $q\approx p$, with
100x original coverage of the target. Note that original coverage of the target genome by the read set is $N/G$.
For the three datasets in our experiments (see Table~\ref{tab:data}), with coverages 200x, 208x, and 129x, the 
expected coverage $E(C_{\ell \geq k})$ has values 9.12, 9.48, and 5.89, respectively, for our initial 
$k=19$ and for our assumed error rate $p=0.15$.
Hence our adopted threshold $h=4$ is from 0.8 to 1.8 standard deviations 
below the expected coverage
meaning that 
most of the correct $k$-mers should be distinguishable from the erroneous ones.

\subsection{Polishing with multiple alignments}

The error correction performed by LoRDEC$^*$ does not make use of long
range information contained in the reads. In particular, 
approximate repeats of the target are collapsed in the DBG into a path 
with alternative branches. 
In practise such repeat regions are
corrected towards a copy of the repeat but not necessarily towards the correct copy. 
However, the correct copy is more likely uncovered because we choose the path 
that minimises the edit distance between the weak region to be corrected and the 
sequence spelled out by the path. Therefore, if we
have several reads from the same location, the majority of them are
likely corrected towards the correct copy.

Our multiple alignment error correction exploits the long range
similarity of reads by identifying the reads that are likely to originate from the
same genomic location. If the reads contain a repeat area, the most
abundant copy of the repeat present in the reads is likely the correct
one. Then by aligning the reads with each other we can correct them
towards this most abundant copy. The approach we use here bears some
similarity to the method used in Coral~\cite{coral}.

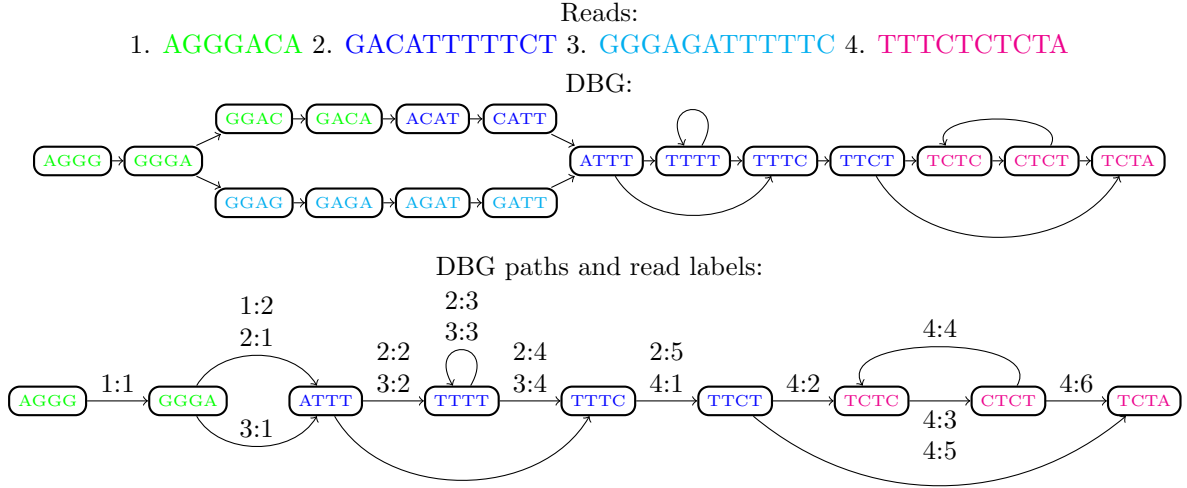
\begin{figure*}
  \centering

{\small
  Reads:\\
  1. {\color{green} AGGGACA} 
  2. {\color{blue} GACATTTTTCT} 
  3. {\color{cyan} GGGAGATTTTTC}
  4. {\color{magenta} TTTCTCTCTA}

  \vspace{0.1 cm}

  DBG:\\
\begin{tikzpicture}[
  kmer/.style={
      rectangle,
      rounded corners,
      text centered,
      thick,
      draw = black,
    }]

    \centering
    \matrix (dbg) [matrix of nodes, column sep=1.5mm, row sep = 1.5mm, nodes={rectangle, rounded corners, text centered, thick, draw = black}]{
    &&\node{\tiny \color{green} GGAC}; & \node{\tiny \color{green} GACA}; & \node{\tiny \color{blue} ACAT}; & \node{\tiny \color{blue} CATT}; \\
    \node{\tiny \color{green} AGGG}; & \node{\tiny \color{green} GGGA}; &&&&&\node{\tiny \color{blue} ATTT}; & \node{\tiny \color{blue} TTTT}; & \node{\tiny \color{blue} TTTC}; & \node{\tiny \color{blue} TTCT}; & \node{\tiny \color{magenta} TCTC}; & \node{\tiny \color{magenta} CTCT}; &\node{\tiny \color{magenta} TCTA};\\
    &&\node{\tiny \color{cyan} GGAG}; & \node{\tiny \color{cyan} GAGA}; & \node{\tiny \color{cyan} AGAT}; & \node{\tiny \color{cyan} GATT}; \\
    };

\draw[->] (dbg-2-1) -- (dbg-2-2);
\draw[->] (dbg-2-2) -- (dbg-1-3);
\draw[->] (dbg-2-2) -- (dbg-3-3);
\draw[->] (dbg-1-3) -- (dbg-1-4);
\draw[->] (dbg-1-4) -- (dbg-1-5);
\draw[->] (dbg-1-5) -- (dbg-1-6);
\draw[->] (dbg-1-6) -- (dbg-2-7);
\draw[->] (dbg-3-3) -- (dbg-3-4);
\draw[->] (dbg-3-4) -- (dbg-3-5);
\draw[->] (dbg-3-5) -- (dbg-3-6);
\draw[->] (dbg-3-6) -- (dbg-2-7);
\draw[->] (dbg-2-7) -- (dbg-2-8);
\draw[->] (dbg-2-8) -- (dbg-2-9);
\draw[->] (dbg-2-9) -- (dbg-2-10);
\draw[->] (dbg-2-10) -- (dbg-2-11);
\draw[->] (dbg-2-11) -- (dbg-2-12);
\draw[->] (dbg-2-12) -- (dbg-2-13);

\draw[->] (dbg-2-8) to [out=60, in=120, looseness=8] (dbg-2-8);
\draw[->] (dbg-2-7) to [out=300, in=240, looseness=1] (dbg-2-9);
\draw[->] (dbg-2-10) to [out=300, in=240, looseness=1] (dbg-2-13);
\draw[->] (dbg-2-12) to [out=60, in=120, looseness=1] (dbg-2-11);
\end{tikzpicture}

\vspace{-0.2cm}
DBG paths and read labels:

\begin{tikzpicture}[
  kmer/.style={
      rectangle,
      rounded corners,
      text centered,
      thick,
      draw = black,
    }]
\matrix (dbg) [matrix of nodes, column sep=8mm, row sep = 3mm, nodes={rectangle, rounded corners, text centered, thick, draw = black}]{
\node{\tiny \color{green} AGGG}; & \node{\tiny \color{green} GGGA}; &\node{\tiny \color{blue} ATTT}; & \node{\tiny \color{blue} TTTT}; & \node{\tiny \color{blue} TTTC}; & \node{\tiny \color{blue} TTCT}; & \node{\tiny \color{magenta} TCTC}; & \node{\tiny \color{magenta} CTCT}; &\node{\tiny \color{magenta} TCTA};\\
};

\draw[->] (dbg-1-1) -- node [above] {1:1} (dbg-1-2);
\draw[->] (dbg-1-2) to [out=60, in=120] node[above,align=left] {1:2\\2:1}   (dbg-1-3);
\draw[->] (dbg-1-2) to [out=300, in=240] node[above] {3:1} (dbg-1-3);
\draw[->] (dbg-1-3) -- node[above,align=left] {2:2\\3:2} (dbg-1-4);
\draw[->] (dbg-1-4) -- node[above,align=left] {2:4\\3:4} (dbg-1-5);
\draw[->] (dbg-1-5) -- node[above,align=left] {2:5\\4:1} (dbg-1-6);
\draw[->] (dbg-1-6) -- node[above] {4:2} (dbg-1-7);
\draw[->] (dbg-1-7) -- node[below,align=left] {4:3\\4:5} (dbg-1-8);
\draw[->] (dbg-1-8) -- node[above] {4:6} (dbg-1-9);

\draw[->] (dbg-1-4) to [out=60, in =120, looseness=8] node[above,align=left] {2:3\\3:3} (dbg-1-4);
\draw[->] (dbg-1-3) to [out=300, in =240, looseness=1] (dbg-1-5);
\draw[->] (dbg-1-6) to [out=320, in =220, looseness=1] (dbg-1-9);
\draw[->] (dbg-1-8) to [out=60, in =120, looseness=1] node[above] {4:4} (dbg-1-7);

\end{tikzpicture}
}

\caption{Augmented DBG. For simplicity reverse complements are not considered. 
The lower graph only shows the branching nodes of the DBG and the labels on the paths/edges are of 
the form {\em read id}:{\em read part id}. 
For example, the path for read 2 consists of segments with labels 2:1, 2:2, 2:3, 2:4, and 2:5.}
\label{fig:dbg}
\end{figure*}

As preprocessing phase for the method, we build a DBG of all the reads using abundancy
threshold $h=1$ to ensure that all $k$-mers present in the reads are
indexed. Then we enumerate the simple paths of the DBG and find for each read the unique path 
that spells it out. 
Each such path is composed of non-overlapping unitig segments that have no branches. We call
such segments the parts of a path.
We associate
to each path segment (i.e., a unitig path of the DBG) a set of triples describing the reads traversing
that segment. Each triple consists of read id, part id, and the direction
of the read on this path. 
Hence the path for a read $i$ consists of segments who have a triplet with $i$ as the read id and 
with part id values 1, 2, ..., 
the path being composed of these segments in the order of the part id value (see Fig.~\ref{fig:dbg}). 
Using this information it is now possible to reconstruct each read from the
DBG except that the reads will be prefixed (suffixed) by the complete
simple path that starts (ends) the read.

In the second phase of our method we take the reads one by one and use
the DBG to select reads that are similar to the current read. We
follow the path for the current read and gather the set of reads
sharing $k$-mers with it, which can be done using the triplets of the augmented DBG. 
Out of these reads we then first select each read $R$ such that the shared $k$-mers span at least 80\% of the shorter one of the
read $R$ and the current read. Furthermore, out of these reads we select those that
share the most $k$-mers with the current read. We call this read set
the {\em friends} of the current read. The number of selected reads is a
parameter of our method (by default 7).

We then proceed to compute a multiple alignment of the current read and its friends. 
To keep the running time feasible, we use the same simple method as
in Coral~\cite{coral}. 
First the current
read is set to be the initial consensus. Then we take each friend of the current read one by
one, align them against the current consensus using banded alignment,
and finally update the consensus according to the alignment. Finally we inspect every column of the multiple
alignment and correct the current read towards the consensus if the
consensus is supported by at least two reads.

We implemented the above procedure in a tool called LoRMA (Long Read Multiple Alignments) using the GATB library~\cite{gatb} for the implementation of the DBG.

\section{Experimental results}

We ran experiments on three data sets that are detailed in Table~\ref{tab:data}. 
The simulated {\em E. coli} data set was generated with PBSIM~\cite{pbsim} using the following parameters: 
mean accuracy 85\%, average read length 10,000, and minimum read length 1,000. 
The other two data sets are real data.
Although our method works solely on the PacBio reads, the table also includes statistics of complementary Illumina reads 
that were used to compare our method against hybrid methods that need also short reads. 
All experiments were run on 32~GB RAM machines equipped with 8~cores.


\begin{table*}
  \centering
  \caption{Data sets used in the experiments}
  \label{tab:data}

{\footnotesize
  \begin{tabular}{lccc}
    \hline
     & {\bf {\em E. coli} (simulated)} &  {\bf {\em E. coli}} & {\bf Yeast} \\
    \hline
    {\bf Reference organism}& & \\
    Name & {\em Escherichia coli} & {\em Escherichia coli} & {\em Saccharomyces cerevisiae} \\
    Strain & K-12 substr. MG1655& K-12 substr. MG1655 & W303 \\
    Reference sequence & NC\_000913& NC\_000913 & CM001806-CM001823 \\
    Genome size & 4.6 Mbp & 4.6 Mbp & 12 Mbp \\
    \hline
    \textbf{PacBio data} &&\\
    Number of reads & 92818 & 89481 & 261964 \\
    Avg. read length & 9997 & 10779 & 5891 \\
    Coverage & 200x & 208x & 129x \\
    \hline
    \textbf{Illumina data} &&\\
    Accession number &- & ERR022075 & SRR567755 \\
    Number of reads &-& 2316613 & 4503422 \\
    Read length &-& 100 & 100 \\
    Coverage  &-& 50x & 38x \\
    \hline
  \end{tabular}
}
\end{table*}

\subsection{Evaluation of the quality of error correction}

In the simulated data set the genomic position where each read derives from is known. Therefore the quality of error correction on the simulated data set is evaluated by aligning the corrected read against the corresponding correct genomic sequence. We allow free deletions in the flanks of the corrected read because the tools trim regions they are not able to correct. To check if the corrected reads align to the correct genomic position, we aligned the corrected reads on the reference genome with BLASR~\cite{blasr} keeping only a single best alignment for each read. The following statistics were computed:
\begin{itemize}
\item {\bf Size}: The relative size of the corrected read set as compared to the original one.
\item {\bf Error rate}: The number of substitutions, insertions and deletions divided by the length of the correct genomic sequence.
\item {\bf Correctly aligned}: The relative number of reads that align to the same genomic position where the read derives from.
\end{itemize}

To evaluate the quality of error correction on the real data sets, we used BLASR~\cite{blasr} to align the original and corrected reads on the reference genome. For each read we used only a single best alignment because a correct read should only have one continuous alignment against the reference. Thus chimeric reads will be only partially aligned.
We computed the following statistics:
\begin{itemize}
\item {\bf Size}: The relative size of the corrected read set as compared to the original one.
\item {\bf Aligned}: The relative size of the aligned regions as compared to the complete read set.
\item {\bf Error rate}: The number of substitutions, insertions and deletions in the aligned regions divided by the length of the aligned regions in the reference sequence.
\item {\bf Genome coverage}: The proportion of the genome covered by the aligned regions of the reads.
\end{itemize}
Together these statistics measure three aspects of the quality of error correction. Size measures the throughput of the method. Aligned and error rate together measure the accuracy of correction. Finally genome coverage estimates if reads deriving from all regions of the genome are corrected.

\subsection{Parameters of our method}

We ran experiments on the real {\em E. coli} data set to test the effect of parameters on the performance of our method. First we tried several progressions of $k$ in the first phase where LoRDEC$^*$ is run iteratively. We started all iterations with $k=19$ because given the high error rate of the data $k$ must be small for correct $k$-mers to occur in the read data. The results of these experiments are presented in Table~\ref{tab:kprog}. With more iterations the size of the corrected read set and the aligned proportion of reads decrease, but the aligned regions are more accurate. The decrease in the size of the corrected read set may be a result of better correction because PacBio reads have more insertions than deletions. However, the decrease in the aligned proportion of the reads may indicate some accumulation of false corrections. The runtime of the method increases with the number of iterations but later iterations take less time as the reads have already been partially corrected during the previous rounds. To balance out these effects, we chose to use a moderate number of iterations, i.e.\ $k=19,40,61$, by default which also optimises the error rate of the aligned regions.

\begin{table*}
  \centering
  \caption{The progression of $k$ for the iterations of LoRDEC$^*$. All results are shown for the whole correction process LoRDEC$^*$+LoRMA.}
  \label{tab:kprog}
{\footnotesize
    \begin{tabular}{lrrrr}
      \hline
      {\bf $k$ progression} & {\bf Size (\%)} & {\bf Aligned (\%)} & {\bf Error rate (\%)} & {\bf Elapsed time (h)} \\
      \hline
      19 & 64.901 & 99.499 & 0.294 & 4.08 \\
      \hline
      19,22,25,28,31 & 66.702 & 99.302 & 0.276 & 12.97 \\
      19,22,25,28,31,34,37,40,43,46 & 66.630 & 99.311 & 0.274 & 20.65 \\
      19,22,25,28,31,34,37,40,43,46,49,52,55,58,61 & 66.546 & 99.296 & 0.271 & 27.53 \\
      \hline
      19,26,33 & 66.401 & 99.329 & 0.274 & 9.58 \\
      19,26,33,40,47 & 66.230 & 99.298 & 0.271 & 13.07 \\
      19,26,33,40,47,54,61 & 66.144 & 99.283 & 0.266 & 16.08 \\
      \hline
      19,33 & 66.705 & 99.358 & 0.277 & 7.68 \\
      19,33,47 & 66.178 & 99.352 & 0.268 & 10.58 \\
      19,33,47,61 & 65.991 & 99.301 & 0.261 & 11.92 \\
      \hline
      19,40 & 66.619 & 99.360 & 0.272 & 8.32 \\
      19,40,61 & 66.223 & 99.317 & 0.257 & 10.30 \\
      \hline
    \end{tabular}
}
\end{table*}

LoRMA also builds a DBG of the reads and thus we need to specify $k$. We investigated the effect of the value of $k$ on the {\em E. coli} data set. Table~\ref{tab:lormak} shows the effect of $k$ on the performance of LoRMA. Because the DBG is only used to detect similar reads in LoRMA, the performance is not greatly affected by the choice of $k$. There is a slight decrease in the throughput of the method as $k$ increases as well as a slight increase in runtime but these effects are very modest. For the rest of the experiments we set $k=19$.

\begin{table}
  \centering
  \caption{The effect of the $k$-mer size in LoRMA. All results are shown for the whole correction process LoRDEC$^*$+LoRMA.}
  \label{tab:lormak}
  {\footnotesize
    \begin{tabular}{lrrrrrr}
      \hline
      {\bf $k$} & {\bf Size} & {\bf Aligned} & {\bf Error rate} & {\bf Elapsed time} & {\bf Memory peak} \\
      & {\bf (\%)} & {\bf (\%)} & {\bf (\%)} & {\bf (h)} & {\bf (GB)} \\
      \hline
      19 & 66.238 & 99.306 & 0.256 & 10.38 & 17.197 \\
      40 & 66.170 & 99.309 & 0.258 & 10.53 & 16.958 \\
      61 & 65.941 & 99.313 & 0.261 & 13.87 & 16.908 \\
      \hline
    \end{tabular}
}
\end{table}

Another parameter of the method is the size of the set of friends of the current read (-friends parameter). We tested also the effect of this parameter on the {\em E. coli} data set. As the optimal value of this parameter might depend on the coverage of the data set, we created several subsets of this data set with different coverage to investigate this. Table~\ref{tab:friends} shows the results of these experiments. We can see that the accuracy of the correction increases as the size of the friends set increases. However, for the data set with the lowest coverage, 75x, the coverage of the genome by the corrected reads decreases when the size of the friends set is increased indicating that lower coverage areas are not well corrected. We can also see that increasing the size of the friends set increases the running time of the method. In the interest of keeping the running time reasonable, we decided to set the default value of the parameter at a fairly low value, 7.

\begin{table}[t]
  \centering
  \caption{The effect of the size of the friends set on the quality of the correction. All results are shown for the whole correction process LoRDEC$^*$+LoRMA.}
  \label{tab:friends}
{\footnotesize
    \begin{tabulary}{1.0\textwidth}{LRRRRR}
      \multicolumn{6}{c}{\bf Coverage 75x}\\
      \hline
      {\bf Friends} & {\bf 5} & {\bf 7} & {\bf 10} & {\bf 15} & {\bf 20} \\
      \hline
      {\bf Size (\%)} & 59.173 & 59.164 & 59.146 & 59.109 & 59.085 \\
      {\bf Aligned (\%)} & 98.894 & 98.983 & 99.099 & 99.192 & 99.226 \\
      {\bf Error rate (\%)} & 0.169 & 0.156 & 0.148 & 0.131 & 0.128 \\
      {\bf Gen. cov. (\%)} & 90.918 & 90.907 & 90.900 & 90.888 & 90.884 \\
      {\bf Elapsed time (h)} & 1.13 & 1.22 & 1.53 & 1.88 & 2.27 \\
      {\bf Memory (GB)} & 14.522 & 14.518 & 14.522 & 14.515 & 14.525 \\
      {\bf Disk (GB)} & 1.076 & 1.076 & 1.076 & 1.076 & 1.076 \\
      \hline
      \\
      \multicolumn{6}{c}{\bf Coverage 100x}\\
      \hline
      {\bf Friends} & {\bf 5} & {\bf 7} & {\bf 10} & {\bf 15} & {\bf 20} \\
      \hline
      {\bf Size (\%)} & 65.759 & 65.738 & 65.723 & 65.670 & 65.607 \\
      {\bf Aligned (\%)} & 98.091 & 98.317 & 98.491 & 98.556 & 98.620 \\
      {\bf Error rate(\%)} & 0.152 & 0.140 & 0.134 & 0.114 & 0.110 \\
      {\bf Gen. cov. (\%)} & 99.404 & 99.403 & 99.405 & 99.403 & 99.405 \\
      {\bf Elapsed time (h)} & 2.53 & 3.32 & 4.32 & 5.80 & 7.08 \\
      {\bf Memory (GB)} & 14.720 & 14.720 & 14.712 & 14.723 & 14.720 \\
      {\bf Disk (GB)} & 1.417 & 1.416 & 1.417 & 1.416 & 1.416 \\
      \hline
      \\
      \multicolumn{6}{c}{\bf Coverage 175x} \\
      \hline
      {\bf Friends} & {\bf 5} & {\bf 7} & {\bf 10} & {\bf 15} & {\bf 20} \\
      \hline
      {\bf Size (\%)} & 66.933 & 66.906 & 66.905 & 66.852 & 66.816 \\
      {\bf Aligned (\%)} & 98.927 & 98.973 & 99.153 & 99.011 & 99.104 \\
      {\bf Error rate(\%)} & 0.222 & 0.194 & 0.191 & 0.140 & 0.133 \\
      {\bf Gen. cov. (\%)} & 100.000 & 100.000 & 100.000 & 100.000 & 100.000 \\
      {\bf Elapsed time (h)} & 6.77 & 8.35 & 10.62 & 14.07 & 17.22 \\
      {\bf Memory (GB)} & 16.009 & 16.016 & 16.003 & 16.002 & 16.006 \\
      {\bf Disk (GB)} & 2.361 & 2.361 & 2.362 & 2.362 & 2.362 \\
      \hline
    \end{tabulary}
}
\end{table}

\subsection{Comparison against previous methods}

We compared our new method against PBcR~\cite{pacbiotoca,mhap} which
is to the best of our knowledge the only previous selfcorrection method for long reads, and LoRDEC~\cite{lordec},
proovread~\cite{proovread} and Jabba~\cite{jabba} which also use short complementary reads. 
Table~\ref{tab:simulated} shows the results on the simulated data set comparing our new method to PBcR using long reads only.
Table~\ref{tab:comparison} shows the results of the comparison of our new method against previous methods on the real data sets. In the following we will use LoRDEC to refer to the hybrid correction method using also short reads and LoRDEC$^*$+LoRMA for our new method in which LoRDEC$^*$ is run in long reads selfcorrection mode followed by LoRMA.

PBcR pipeline from Celera Assembler version 8.3rc2 was run without the assembly phase and memory limited to
16 GB. PBcR was run both only using PacBio reads and by utilising also the short read data. For PBcR utilising also short read data,  the PacBio reads were divided into three subsets each of which was corrected in its own run.  Proovread v2.12 was run with the sequence/fastq files chunked to 20M
as per the usage manual and used 16 mapping threads. LoRDEC used an
abundancy threshold of 3 and $k$-mer size was set to 19 similar to the experiments by Salmela and Rivals~\cite{lordec}. Jabba 1.1.0 used $k$-mer size 31 and short output mode. LoRMA was run
with 6 threads. The $k$-mer sizes for LoRDEC$^*$+LoRMA iteration steps were chosen
19, 40 and 61. For proovread and LoRDEC we present results for trimmed and split reads.

Table~\ref{tab:simulated} shows that on the simulated data both PBcR and LoRDEC$^*$+LoRMA are able to correct most of the data. Our new method achieves a lower error rate and higher throughput. We see that the fraction of corrected reads aligning to the correct genomic position is lower for LoRDEC$^*$+LoRMA than for PBcR when all reads are considered, which suggests that LoRDEC$^*$+LoRMA tends to overcorrect some reads. However, for corrected reads longer than 2000 bp this difference disappears and thus we can conclude that the overcorrected reads are short.
When compared to the other selfcorrection method, PBcR, our new tool has a higher throughput and produces more accurate results on both real data sets as shown in Table~\ref{tab:comparison}. Out of the hybrid methods, Jabba has a lower error rate than LoRDEC$^*$+LoRMA but its throughput is lower. When compared to the other hybrid methods, LoRDEC$^*$+LoRMA has comparable accuracy and throughput. All hybrid methods produce corrected reads that do not cover the whole {\em E. coli} reference, which could be a result of coverage bias in the Illumina data. On the yeast data proovread produced few corrected reads and thus the coverage of the corrected reads is very low. 

Table~\ref{tab:comparison} shows that our method is slower and uses more memory than PBcR in selfcorrection mode but its disk usage is lower. On the {\em E. coli} data set our new method is faster than proovread and PBcR utilising short read data but slower than LoRDEC, Jabba or PBcR using only PacBio data. On the yeast data set we are faster than PBcR in hybrid mode but slower than the others. 

On the {\em E. coli} and yeast data sets, LoRDEC$^*$+LoRMA uses 45\% and 37\%, respectively, of its running time on LoRDEC$^*$ iterations. On both data sets the error rate of the reads after LoRDEC$^*$ iterations and trimming was 0.5\%.

\begin{table*}
  \centering
  \caption{Comparison of LoRDEC$^*$+LoRMA against PBcR (PacBio only) on the simulated {\em E. coli} data set}
  \label{tab:simulated}
{\footnotesize
  \begin{tabular}{lrrrrrrr}
    \hline
    {\bf Tool} & {\bf Size} & {\bf Error} & {\bf Correctly} & {\bf Correctly aligned} & {\bf Elapsed} & {\bf Memory} & {\bf Disk} \\
               &            & {\bf rate}  & {\bf aligned}   & {\bf  $\geq$ 2000 bp}   & {\bf time}    & {\bf peak} & {\bf peak} \\
    & {\bf (\%)} & {\bf (\%)} & {\bf (\%)} &{\bf (\%)} & {\bf (h)} &{\bf (GB)} & {\bf (GB)} \\
    \hline
    Original & 100.000 & 13.015 & 99.997 & 99.997 & - & - & - \\
    PBcR (PacBio only) & 92.457 & 0.604 & 99.953 & 99.984 &2.63 & 9.066 & 17.823 \\
    LoRDEC$^*+$LoRMA & 94.372 & 0.109 & 96.866 & 99.987 &14.30 & 17.338 & 3.192 \\
    \hline
  \end{tabular}
}
\end{table*}

\begin{table*}
  \centering
  \caption{Comparison of both hybrid and selfcorrection tools on PacBio data. Results for tools utilising also Illumina data are shown on a grey background.}
  \label{tab:comparison}
{\footnotesize
  \begin{tabu}{llrrrrrrr}
    \hline
    & {\bf Tool} & {\bf Size} & {\bf Aligned} & {\bf Error rate} & {\bf Genome}   & {\bf Elapsed } & {\bf Memory} & {\bf Disk}\\
    &            &            &               &                  & {\bf coverage} & {\bf time}     & {\bf peak}   & {\bf peak}\\
    & & {\bf (\%)} & {\bf (\%)} & {\bf (\%)} & {\bf (\%)} & {\bf (h)} & {\bf (GB)} & {\bf (GB)} \\
     \hline
     \multirow{6}{*}{\rotatebox[origin=c]{90}{\bf E. coli}}
      & Original & 100.000 & 71.108 & 16.9126 & 100.000 & - & - & -\\
      & \cellcolor{black!20}LoRDEC & \cellcolor{black!20} 65.672 & \cellcolor{black!20} 98.944 & \cellcolor{black!20} 0.1143 & \cellcolor{black!20} 99.820 & \cellcolor{black!20} 0.96 & \cellcolor{black!20} 0.368 & \cellcolor{black!20} 1.570 \\
      & \cellcolor{black!20}proovread & \cellcolor{black!20} 61.590 & \cellcolor{black!20} 98.603 & \cellcolor{black!20} 0.2789 & \cellcolor{black!20} 99.728 & \cellcolor{black!20} 28.65 & \cellcolor{black!20} 9.522 & \cellcolor{black!20} 7.174 \\
      & \cellcolor{black!20}PBcR (with Illumina) & \cellcolor{black!20} 52.103 & \cellcolor{black!20} 98.507 & \cellcolor{black!20} 0.0682 & \cellcolor{black!20} 98.769 & \cellcolor{black!20} 15.13 & \cellcolor{black!20} 17.429 & \cellcolor{black!20} 160.154 \\
      & \cellcolor{black!20}Jabba & \cellcolor{black!20} 2.873 & \cellcolor{black!20} 99.945 & \cellcolor{black!20} 0.0003 & \cellcolor{black!20} 99.745 & \cellcolor{black!20} 0.02 & \cellcolor{black!20} 0.168 & \cellcolor{black!20} 0.606 \\
      & PBcR (only PacBio) & 51.068 & 86.023 & 0.6905 & 100.000 & 1.68 & 22.00 & 16.070 \\
      & LoRDEC$^*$+LoRMA & 66.223 & 99.318 & 0.2572 & 100.000 & 10.40 & 16.984 & 2.824 \\
     \hline
     \multirow{6}{*}{\rotatebox[origin=c]{90}{\bf Yeast}}
      & Original & 100.000 & 89.929 & 16.8442 & 99.974 & - & - & - \\
      & \cellcolor{black!20}LoRDEC & \cellcolor{black!20} 75.522 & \cellcolor{black!20} 97.337 & \cellcolor{black!20} 0.9987 & \cellcolor{black!20} 99.833 & \cellcolor{black!20} 3.17 & \cellcolor{black!20} 0.451 & \cellcolor{black!20} 2.776 \\
      & \cellcolor{black!20}proovread & \cellcolor{black!20} 0.306 & \cellcolor{black!20} 97.156 & \cellcolor{black!20} 0.8004 & \cellcolor{black!20} 20.346 & \cellcolor{black!20} 11.18 & \cellcolor{black!20} 4.764 & \cellcolor{black!20} 7.162  \\
      & \cellcolor{black!20}PBcR (with Illumina) & \cellcolor{black!20} 57.337 & \cellcolor{black!20} 98.100 & \cellcolor{black!20} 0.3342 & \cellcolor{black!20} 99.652 & \cellcolor{black!20} 22.05 & \cellcolor{black!20} 20.085 & \cellcolor{black!20} 157.726 \\
      & \cellcolor{black!20}Jabba & \cellcolor{black!20} 24.979 & \cellcolor{black!20} 99.484 & \cellcolor{black!20} 0.1279 & \cellcolor{black!20} 99.900 & \cellcolor{black!20} 0.17 & \cellcolor{black!20} 1.031 & \cellcolor{black!20} 0.993 \\
      & PBcR (only PacBio) & 60.065 & 95.822 & 2.1018 & 99.907 & 4.42 & 9.571 & 24.610 \\
      & LoRDEC$^*$+LoRMA & 71.987 & 98.088 & 0.3644 & 99.375 & 21.08 & 17.968 & 4.852 \\
    \hline
   \end{tabu}
}
\end{table*}

\subsection{The effect of coverage}

Especially for larger genomes it is of interest to know how much
coverage is needed for the error correction to succeed. We investigated
this by creating random subsets of the {\em E. coli} data set
with coverages 25x, 50x, 100x, and 150x. We then ran our method
and PBcR~\cite{pacbiotoca,mhap} on these subsets to investigate the
effect of coverage on the error correction performance.
Table~\ref{tab:coverage} shows the results of these experiments. The
other tools, LoRDEC, Jabba and proovread, use also the complementary Illumina reads and the coverage of PacBio reads does not affect their performance.

When the coverage is high, the new method retains a larger proportion of the reads than PBcR and is more accurate, whereas when the coverage is low, PBcR retains more of the data and a larger proportion of it can be aligned. However, the error rate remains much lower for our new tool. The reads corrected by PBcR also cover a larger part of the reference when the coverage is low.

\begin{table*}
  \caption{The effect of coverage of the PacBio read set on the quality of the correction.}
  \label{tab:coverage}
  \centering
{\footnotesize
    \begin{tabular}{l|rrrrr|rrrrr}
      \hline
       & \multicolumn{5}{c|}{\bf LoRDEC$^*$+LoRMA} & \multicolumn{5}{c}{\bf PBcR}\\
      {\bf Coverage} & {\bf 25x} & {\bf 50x} & {\bf 100x} & {\bf 150x} & {\bf 208x} & {\bf 25x} & {\bf 50x} & {\bf 100x} & {\bf 150x} & {\bf 208x}\\
      \hline
      {\bf Size (\%)} & 3.105 & 30.348 & 65.739 & 67.198 & 66.223 & 31.132 & 44.190 & 48.391 & 50.284 & 51.068 \\
      {\bf Aligned (\%)} & 99.400 & 99.663 & 98.328 & 98.748 & 99.318 & 99.941 & 99.794 & 95.966 & 90.003 & 86.023 \\
      {\bf Error rate (\%)} & 0.329 & 0.187 & 0.140 & 0.159 & 0.257 & 2.224 & 1.396 & 0.874 & 0.757 & 0.6905 \\
      {\bf Gen. cov. (\%)} & 3.886 & 45.763 & 99.403 & 99.999 & 100.000 &94.638 & 100.000 & 100.000 & 100.000 & 100.000\\
      {\bf Time (h)} & 0.10 & 0.32 & 3.30 & 7.17 & 10.40 & 0.08 & 0.18 & 0.47 & 0.93 & 1.68 \\
      {\bf Memory (GB)} & 14.165 & 14.275 & 14.718 & 15.415 & 16.984 & 7.851 & 9.020 & 9.706 & 9.931 & 22.00 \\
      {\bf Disk (GB)} & 0.272 & 0.655 & 1.416 & 2.024 & 2.824 & 1.232 & 2.443 & 3.714 & 7.114 & 16.070\\
      \hline
    \end{tabular}
}
\end{table*}

\section{Conclusions}

We have presented a new method for correcting long and highly erroneous sequencing reads. Our method shows that efficient alignment free methods can be applied to highly erroneous long read data. The current approach needs alignments to take into account the global context of errors. Reads corrected by the new method have an error rate less than half of the error rate of reads corrected by previous selfcorrection methods.
Furthermore, the throughput of the new method is 20\% higher than previous selfcorrection methods with read sets having coverage at least 75x.

Recently several algorithms for updating the DBG instead of constructing it from scratch when $k$ changes have been proposed~\cite{bobo15,cale14}. However, these methods are not directly applicable to our method because also the read set changes when we run LoRDEC$^*$ iteratively on the long reads.

Our method works solely on the long reads, whereas many previous methods require also short accurate reads produced by e.g.~Illumina sequencing, which can incorporate sequencing biases in PacBio reads. This could have very negative effect on sequence quality, especially since Illumina suffers from GC content bias and some context dependent errors~\cite{sc15,na11}.

As further work we plan to improve the method to scale up to mammalian size genomes. We will investigate a more compact representation of the path labels in the augmented DBG to replace the simple hash tables currently used. Construction of multiple alignment could also be improved by exploiting partial order alignments~\cite{Lee2002} which have been shown to work well with PacBio reads~\cite{Chin2013}. 

Another direction of further work is to investigate the applicability of the new method on long reads produced by the Oxford NanoPore MinION platform. Laver et al.~\cite{la15} have reported an error rate of 38.2\% for this platform and they also observed some GC content bias. Both of these factors make the error correction problem more challenging and therefore it will be interesting to see a comparison of the methods on this data.

\section*{Funding} This work was supported by the Academy of Finland (grant 267591 to L.S.), ANR Colib'read (grant ANR-12-BS02-0008), IBC (ANR-11-BINF-0002), and D\'{e}fi MASTODONS to E.R., and EU FP7 project SYSCOL (grant UE7-SYSCOL-258236 to E.U.).

\bibliographystyle{plain}
\bibliography{lr-correction}

\end{document}